\documentclass[twocolumn,showpacs,superscriptaddress,amsmath,amssymb,aps,prl]{revtex4}
\usepackage{graphicx,psfrag}
\usepackage{dcolumn}
\usepackage{bm}
\usepackage[latin1]{inputenc} 
\usepackage{fancyunits}
\begin{document}
\title{Correlations and Synchrony in Threshold Neuron Models}
\author{Tatjana Tchumatchenko}
\affiliation{Max Planck Institute for Dynamics and Self-Organization and Bernstein Center for Computational Neuroscience Göttingen, Germany}
\affiliation{Göttingen Graduate School for Neurosciences and Molecular Biosciences, Germany}
\author{Aleksey Malyshev}
\affiliation{Inst.~of Higher Nervous Activity and Neurophysiology, RAS, Moscow, Russia}
\affiliation{Dep.~of Psychology, University of Connecticut, Storrs, USA}
\author{Theo Geisel}
\affiliation{Max Planck Institute for Dynamics and Self-Organization and Bernstein Center for Computational Neuroscience Göttingen, Germany}
\author{Maxim Volgushev}
\affiliation{Inst.~of Higher Nervous Activity and Neurophysiology, RAS, Moscow, Russia}
\affiliation{Dep.~of Neurophysiology, Ruhr-University Bochum, Germany}
\affiliation{Dep.~of Psychology, University of Connecticut, Storrs, USA}
\author{Fred Wolf}
\affiliation{Max Planck Institute for Dynamics and Self-Organization and Bernstein Center for Computational Neuroscience Göttingen, Germany}
\date{\today}
\begin{abstract}
We study how threshold model neurons transfer temporal and interneuronal input correlations to correlations of spikes. We find that the low common input regime is governed by firing rate dependent spike correlations which are sensitive to the detailed structure of input correlation functions. In the high common input regime the spike correlations are insensitive to the firing rate and exhibit a universal peak shape independent of input correlations. Rate heterogeneous pairs driven by common inputs in general exhibit asymmetric spike correlations.
\end{abstract}
\maketitle
Neurons in the CNS exhibit temporally correlated activity that can reflect specific features of sensory stimuli or behavioral tasks~\cite{ZoharyPsyCor,Lampl}. Recently, the origin, statistical structure and coding properties of spike correlations in neuronal systems have attracted substantial attention~\cite{Mainen1995,Svirskis,Rocha}. How do neurons transfer correlated inputs into correlated output? This fundamental question is vital to understand the structure of network correlations, yet it is unanswered. In the past, most theoretical analyses addressing this question utilized coupled stochastic differential equations and the Fokker Planck formalism~\cite{Rocha,Parga}. These approaches are technically very demanding and are therefore practically restricted to simple stochastic processes, see e.g.~\cite{Parga}. As a result, explicit expressions for quantities of interest are often lacking or obtainable only in special limiting cases.\\
Here we show that an alternative modeling framework, based on the theory of smooth random functions~\cite{JungNoise, RiceNOise}, can provide a mathematically transparent and highly tractable description of spike correlations driven by inputs of arbitrary temporal structure and correlation strength. Our theoretical findings may find applications beyond neuroscience, e.g. in spin ordering, reliability studies, as the statistics of (upward) level crossings is a general, long standing problem in physics and engineering~\cite{RiceNOise}. We calculate quantities which have so far escaped a theoretical description by competing Fokker-Planck based formalism: 1) peak spike correlation for arbitrary input correlation strength, 2) rate independent peak shape in the high correlation regime, 3) complete spike correlation function for weak correlations, 4) asymmetric spike correlation function in rate inhomogeneous pairs. A priori, the simple threshold model we use cannot be expected to completely capture the complex behavior of cortical neurons. Remarkably, our results reproduce and extend previous reports on the firing rate dependence of cortical spike correlations~\cite{Rocha,Svirskis}. Furthermore, we were able to qualitatively confirm all new predictions in our {\em{in vitro}} experiments.\\
\begin{figure}
\begin{minipage}[t]{0.3\linewidth}
\includegraphics[width=\linewidth,clip]{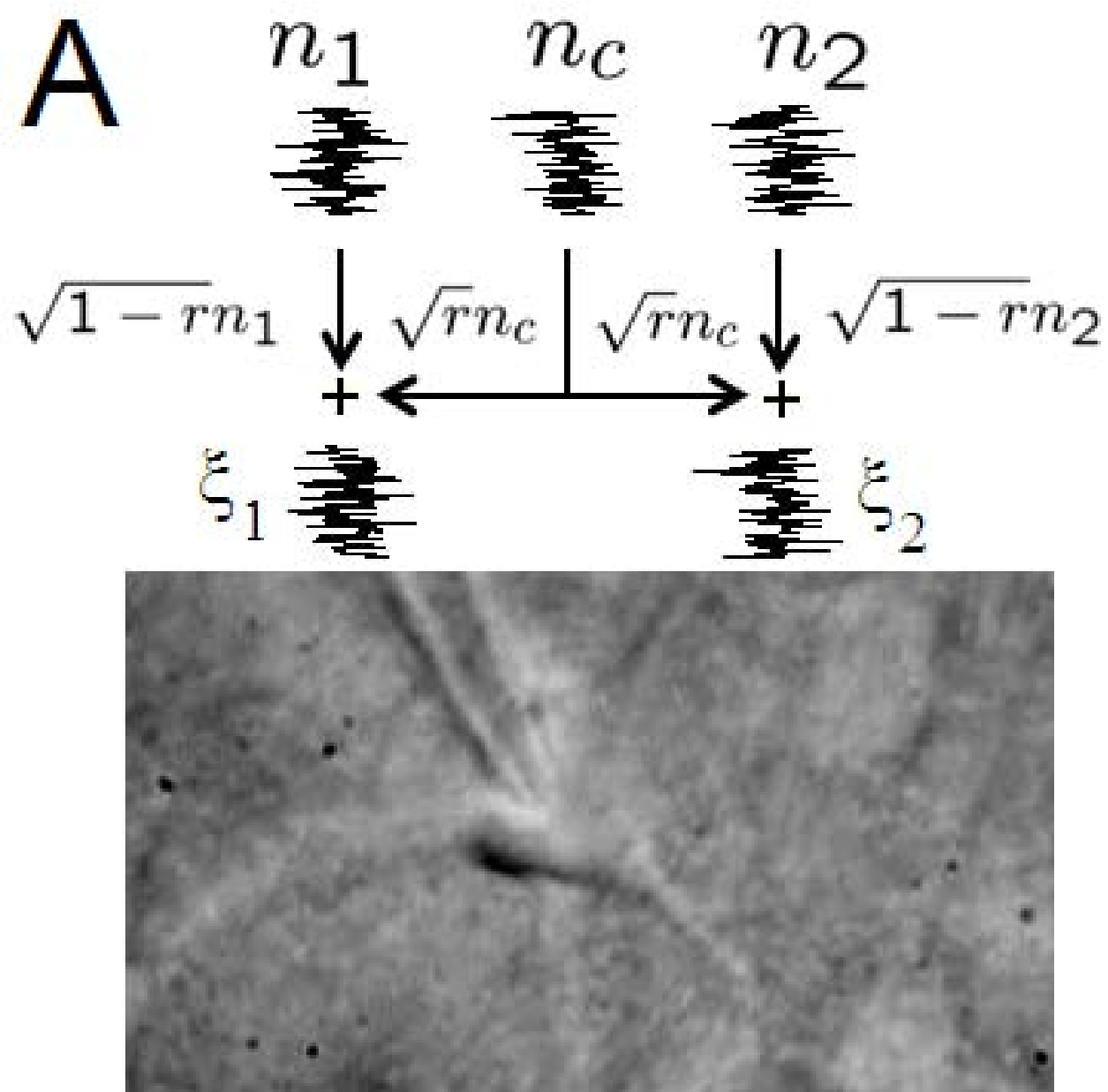}
\end{minipage}
\begin{minipage}[t]{0.65\linewidth}
\includegraphics[width=\linewidth,clip]{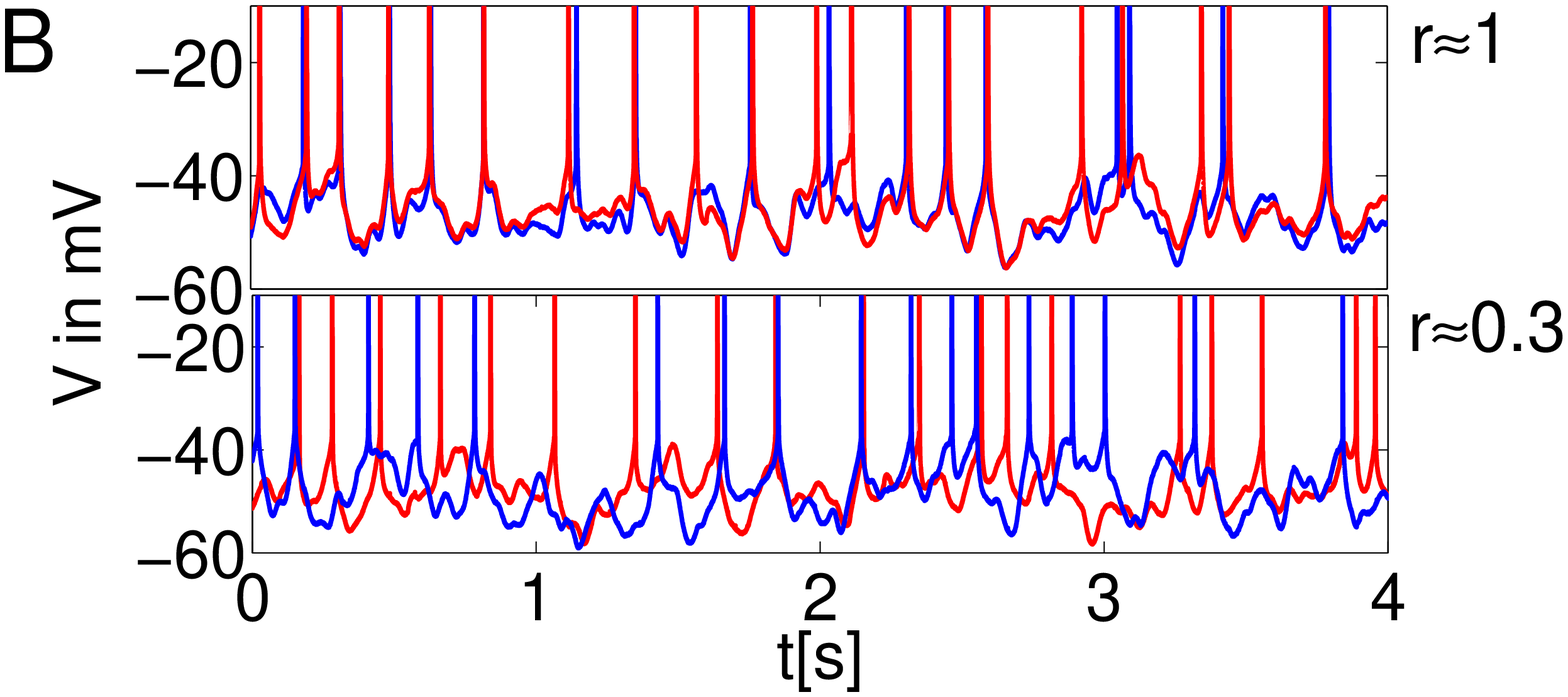}
\end{minipage}
\caption{(A) Paradigm of two fluctuating current traces which share a component $n_C$ and have a correlation strength $r (r>0)$. The correlated currents are injected successively into neurons. (B) MP traces of two neurons (red and blue) recorded in response to correlated fluctuating current with $r\approx 1$ and $r\approx 0.3$. In all four recordings $\nu\approx 5\hertz$, $\tau_s$=$20\milli\second$  and $C_I(\tau)$ as in Eqs.~\ref{CCor},\ref{Cosh}; spikes are truncated at $-10\milli\volt$}
\label{SampleTraces}
\end{figure}
{\em{Framework--}} We model the membrane potential (MP) of a neuron by a temporally continuous, stationary Gaussian random function $V(t)$ with temporal correlation $C(\tau)$$=$$\langle V(0)V(\tau) \rangle$ and zero mean, which we term Gaussian Pseudo Potentials (GPPs); $\langle\cdot\rangle$ denotes the ensemble average. Assuming a simple leaky integrator model with a membrane time constant $\tau_{\text{M}}$, $\tau_{\text{M}}\dot{V}(t)$$=$$-V(t)+\xi(t)$. The voltage and current statistics are linked by:
\begin{align}
\widetilde{C}_I(\omega)&=\widetilde{C}(\omega)(1+\tau_{\text{M}}^2\omega^2).~\label{CCor}
\end{align}
where $C_I(\tau)$ denotes the current correlation function and $\tilde{C}_I(\omega)$ its Fourier transform. To induce correlations we use common input. Two correlated currents $\xi_j(t),~j=1,2$ (1 and 2 denote noise input to neurons 1 and 2) are derived from three statistically independent temporally correlated Gaussian processes $n_1,n_2,n_c$ with current correlation function $C_I(\tau)$:
\begin{gather}
\xi_j(t)=\sqrt{1-r}n_{j}(t)+\sqrt{r}n_c(t)\label{CrossVol}.
\end{gather}
$n_{j}$ are the individual noise components (as in Fig.~\ref{SampleTraces} (A)) and $n_c$ the shared component. $V_j$ and $n_j$ can be related using the membrane filter in Eq.~\ref{CCor}. The corresponding correlated voltages $V_1, V_2$ mimic the neuronal traces of two neurons subject to common input. $r$ modulates between full synchrony and asynchrony, 0$\leq$$r$$<$1.\\
{\em{Model spike statistics--}} The simplest conceivable model of spike generation from a fluctuating MP identifies the spike times with upward threshold crossings of $V(t)$. The spike times $t_j$ are then given by the spike measure
\begin{align}
s(t)&=\sum\delta(t-t_j)=\delta(V(t)-\psi_0)|\dot{V(t)}|\theta(\dot{V}(t)),\label{DefSpikeMeasure}
\end{align}
where $\psi_0$ is the distance to threshold relative to $\langle V(0)\rangle$, $\delta(\cdot)$ and $\theta(\cdot)$ are the Dirac delta and Heaviside theta functions, respectively.
In contrast to the classical integrate-and-fire (IF) model this model has no reset~\cite{Parga} but exhibits an intrinsic silence period after a spike. We assume a smooth $C(\tau)$ such that the variance $\langle\dot{V}(t)\dot{V}(t)\rangle=-C''(0)$ exist and $V(t)$ has a finite rate of threshold crossings. The differential correlation time $\tau_s=\sqrt{C(0)/ |C''(0)|}$ describes the decay of the correlation function near $\tau$$=$$0$. For numerical simulations and experimental testing we choose the following selfcharacteristic, exponentially decaying correlation function: 
\begin{align}
c(\tau)&=1/\cosh(\tau /\tau_s),~~~\langle V(t)V(t+\tau)\rangle=\sigma_V^2c(\tau).\label{Cosh}
\end{align}
However, most results are independent of the particular shape of $C(\tau)$. All spike correlations between two neurons can be obtained via the covariance matrix $C$ and the joint probability density of $\vec{V}$$=$$V_1(0),\dot{V}_1(0),V_2(\tau),\dot{V}_2(\tau)$ $p(\vec{V})$$=$$\exp(-\vec V^TC^{-1}\vec V/2)/(4\pi^2\sqrt{{\rm{Det}}C})$ where
\begin{align}
C=\left(\begin{array}{cccc}\sigma_{V_1}^2& 0&  C_{12}(\tau)&  C'_{12}(\tau)\\
0& \sigma_{\dot{V}_1}^2& -C'_{12}(\tau)& -  C''_{12}(\tau)\\ 
C_{12}(\tau)& -C'_{12}(\tau)& \sigma_{V_2}^2& 0\\
C'_{12}(\tau)& -C''_{12}(\tau)& 0&\sigma_{\dot{V}_2}^2\end{array}\right)\label{DefCorCrossInhom}
\end{align}
Here $C'_{12}(\tau)$ and $C''_{12}(\tau)$ denote the first and second temporal derivative of $C_{1,2}(\tau)$, respectively. Note that Eq.~\ref{DefCorCrossInhom} implies that all pairwise correlations can be expressed as a functional of $C_{12}(\tau),C'_{12}(\tau),C''_{12}(\tau)$. 
Voltage cross correlation function $C_{12}(\tau)$ is:
\begin{align}
C_{12}(\tau)&=\langle V_1(0)V_2(\tau)\rangle=r\sigma_{V_1}\sigma_{V_2} c(\tau).
\end{align}
\begin{figure}[b]
\begin{minipage}[t]{0.48\linewidth}
\includegraphics[width=0.95\linewidth,clip]{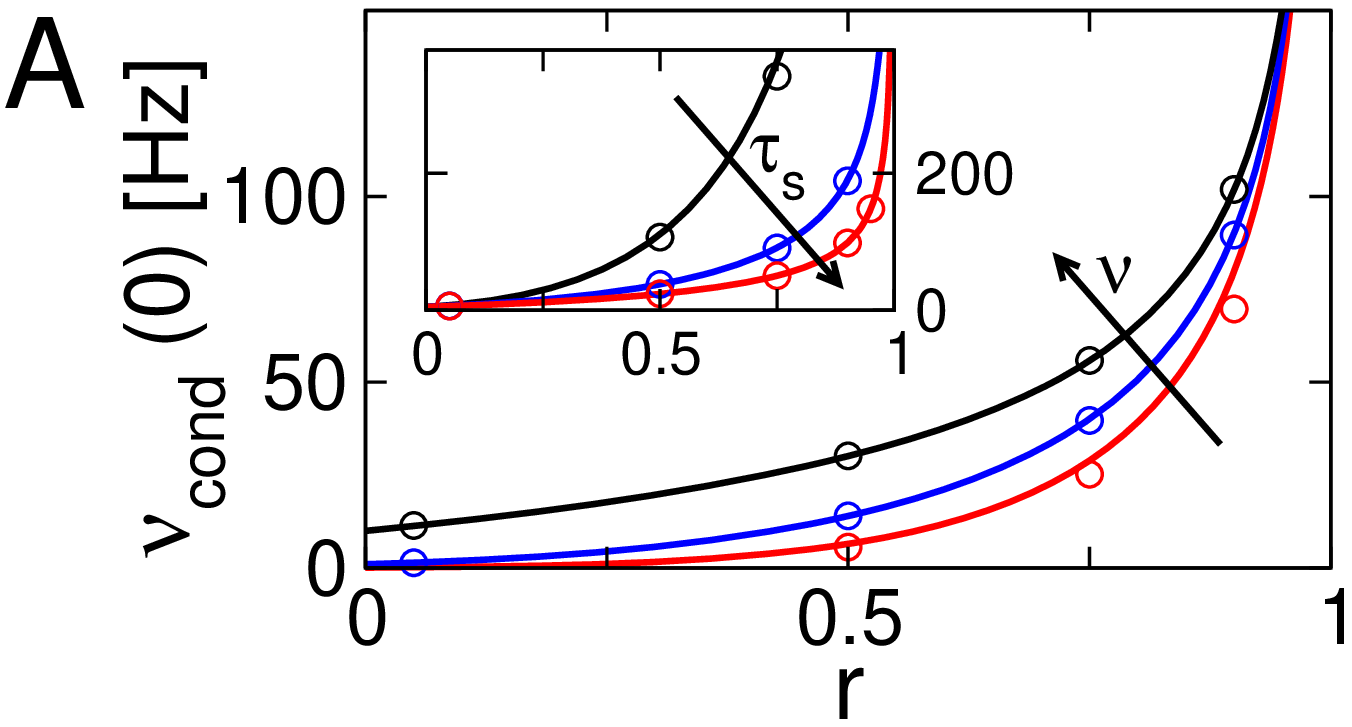}
\end{minipage}
\begin{minipage}[t]{0.48\linewidth}
\centering
\includegraphics[width=\linewidth,clip]{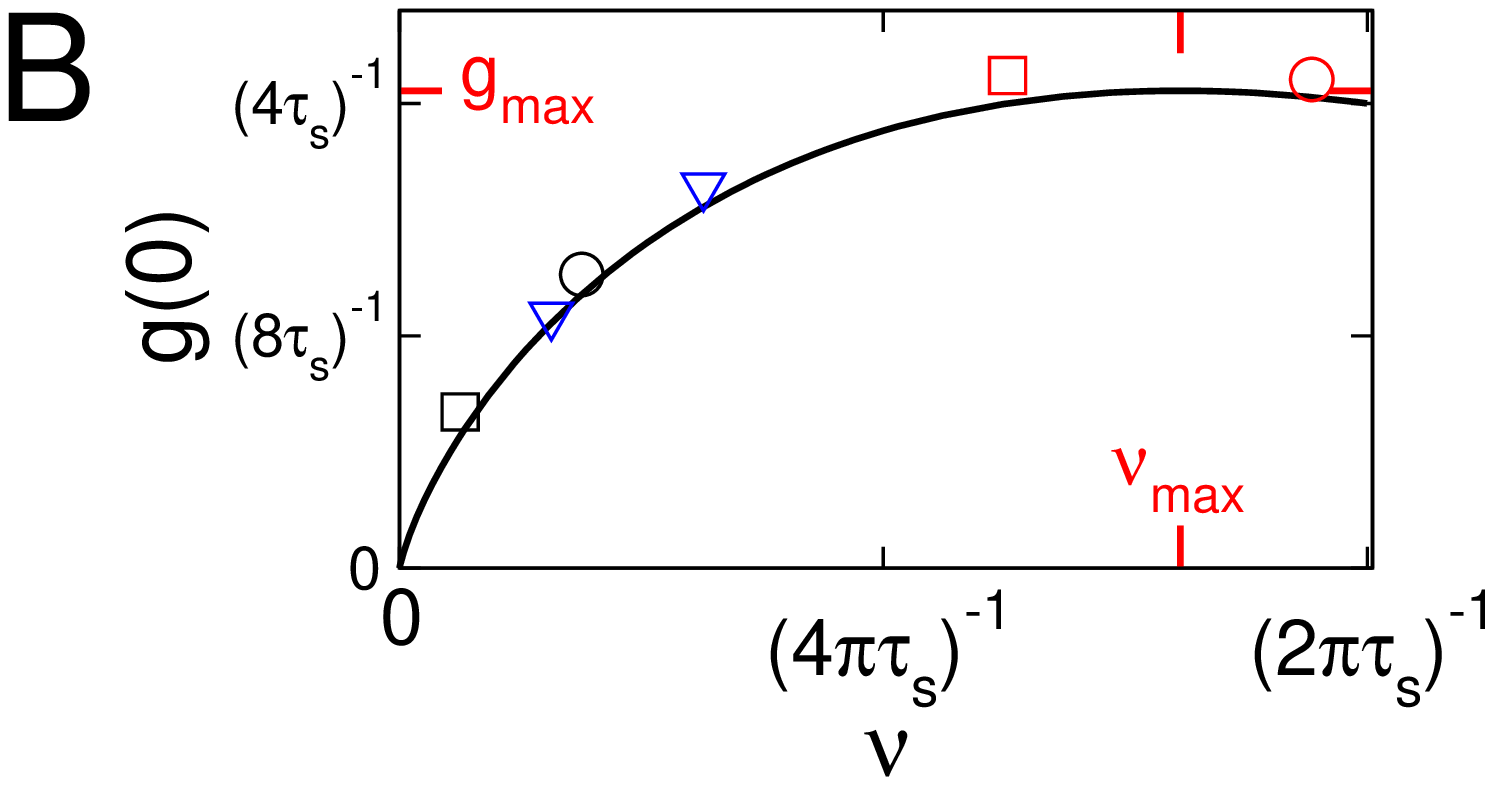}
\end{minipage}
\begin{minipage}[t]{0.48\linewidth}
\includegraphics[width=\linewidth,clip]{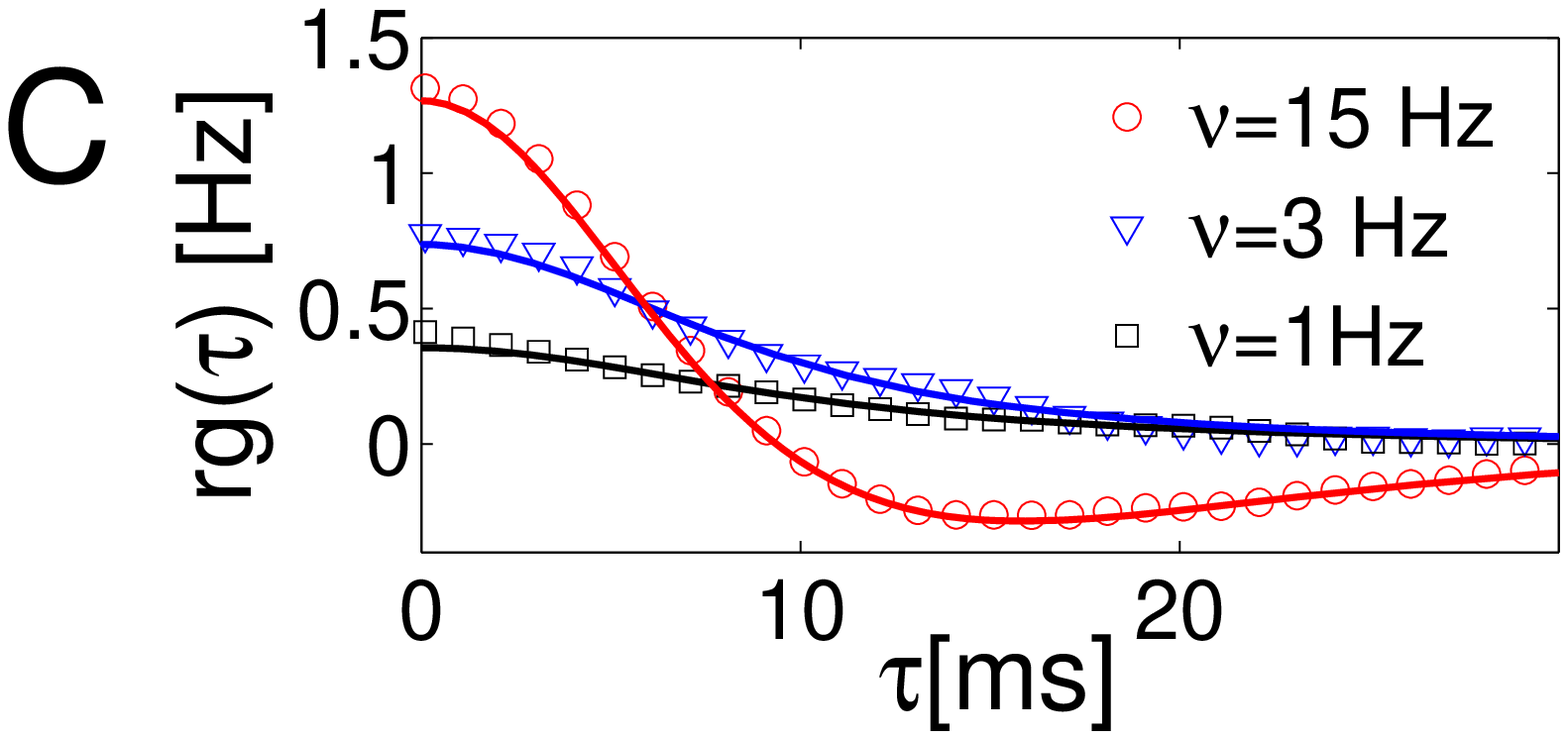}
\end{minipage}
\begin{minipage}[t]{0.48\linewidth}
\includegraphics[width=0.8\linewidth,clip]{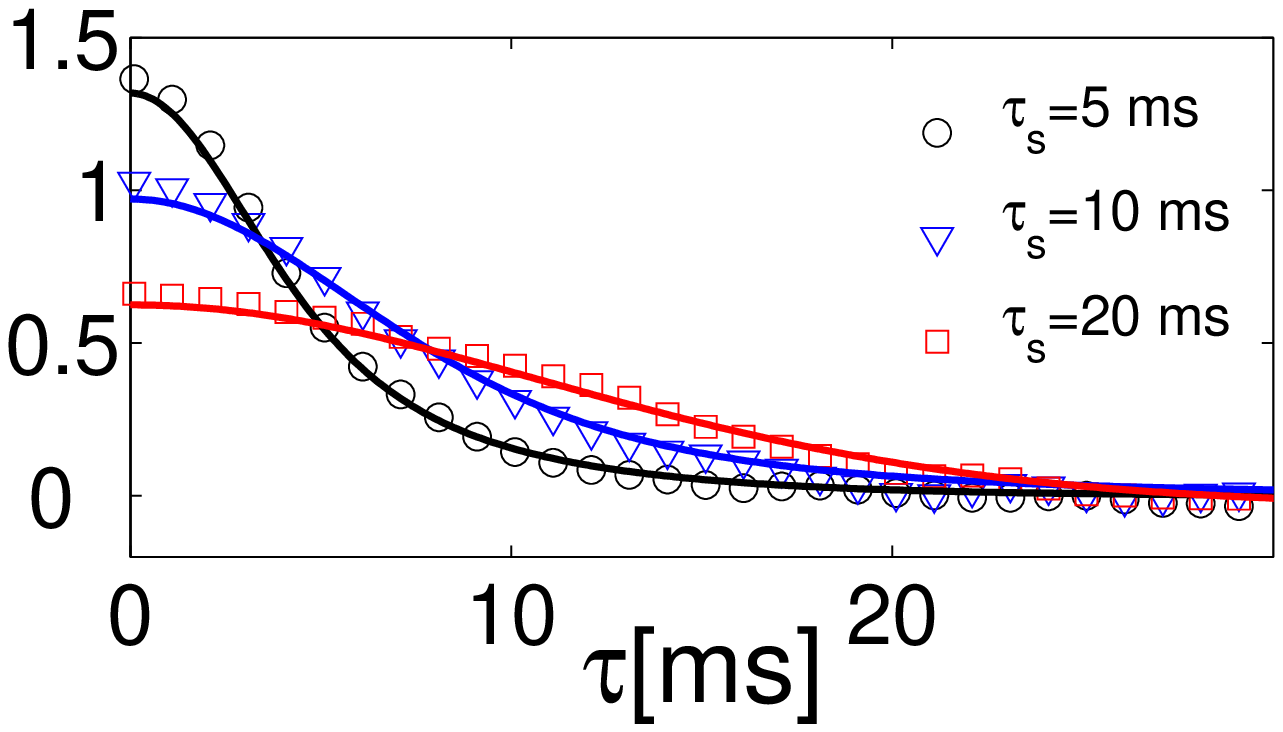}
\end{minipage}
\caption{Simulated and analytical spike cross correlations of two correlated neurons ($\nu_1$=$\nu_2$). (A) $\nu_{cond}(0)$ vs. $r$ as in Eq.~\ref{Synch};  $\nu_{cond}(0)$ for a fixed $\tau_s$=$10\milli\second$ and $\nu$=$0.1\hertz$(red), $\nu$=$1\hertz$(blue), $\nu$=$10\hertz$(black). (inset) $\nu_{cond}(0)$ for a fixed $\nu$=$5\hertz$ and $\tau_s$=$10\milli\second$(red), $\tau_s$=$5\milli\second$(blue), $\tau_s$=$1\milli\second$(black). Circles denote simulated results. (B) $g(0)$ vs. $\nu$ for fixed $\tau_s$ and small $r$. In red $\nu_{max}$ and $g_{max}$, simulation results from (C) are shown as corresponding symbols. (C) $rg(\tau)$ vs. $\tau$ as in Eq.~\ref{SmallRHom} and simulated results for $r$$=$$0.05$ (left) $\nu$=$15,3,1\hertz$, $\tau_s$=$10\milli\second$. (right) $\tau_s$=$5,10,20\milli\second$, $\nu$=$5\hertz$.}\label{RateDepSmallR}
\end{figure}
The firing rate $\langle s(t)\rangle$ of one neuron is then:
\begin{align}
\nu&=\langle s(t)\rangle =\exp(-\psi_0^2/(2\sigma_{V}^2))/(2\pi \tau_s)\label{DefStatRate}
\end{align}
The firing rate in Eq.~\ref{DefStatRate} depends only on two parameters: the correlation time and the threshold-to-variance ratio $\psi_0^2/\sigma_V^2$, but not on the specific choice of the correlation function~\cite{RiceNOise}. Hence, processes with the same $\tau_s$ but different form of $C(\tau)$ will have the same $\nu$, despite different temporal spike statistics. The threshold-to-variance ratio basically determines the probability of threshold crossings. A decrease in $\psi_0/\sigma_V$ leads to an increase in $\nu$. Injections of constant currents shift the mean potential and thus decrease the distance to threshold $\psi_0$ resulting in a higher $\nu$. $\nu$ is also increasing with decreasing $\tau_s$, because faster fluctuations lead to higher rate of threshold crossings. This model has a maximal firing rate $\tilde{\nu}=1/(2\pi\tau_s)$, which corresponds to the upward zero crossings of the random process. It should therefore be used in the fluctuation driven, low firing rate $\nu$$<$$\tilde{\nu}$ regime which is prevalent in cortical neurons~\cite{Greenberg}.\\
{\em{Experimental test--}} To access spike correlations in real neurons {\em{in vitro}}, we made whole-cell recordings from layer 2/3 pyramidal neurons ($N$=$19$) in neocortical slices from rats (PND 22-27). Correlated inputs to neurons were mimicked by injection of digitally synthesized sets of fluctuating currents with $C_I(\tau)$ (Eqs.~\ref{CCor},\ref{Cosh}), varying the correlation parameter ($r\in\{0.3,1\}$), and time constants $\tau_{\text{M}}$=$20\milli\second$, $\tau_s$=$\{20,100\}\milli\second$. Correlated currents were generated by specifying the Fourier spectrum and transferring to the time domain as described in~\cite{RiceNOise}, such that the voltage correlation function $C(\tau)$ of the neurons was similar to the model correlation function (Eq.~\ref{Cosh}). During the recording ($10\second$ or $5\second$ episodes), we targeted two different firing rates $\nu_1$=$5\hertz$ ($T_1$=$10\second$), $\nu_2$=$10\hertz$ ($T_2$=$5\second$), by injection of an additional constant current. The average firing rate obtained is denoted by $\nu_m$.
We obtained a total of $N$=$281$ recordings for $\nu_1$, $N$=$299$ recordings for $\nu_2$. For identical noise injection ($r$=$1$) we recorded $N$=$80$ at the target rate $\nu_1$, $N$=$81$ at $\nu_2$, and $N$=$21$ at a target rate of $3\hertz$ ($T_3$=$T_1$). Fig.~\ref{SampleTraces} shows examples of recorded voltage traces for large and small $r$. We calculated $\nu_{cond}(\tau)$ using a Gaussian filter kernel with $\sigma$=$5\milli\second$ and $95\%$ Jackknife confidence intervalls for $N$ random subsamples each containing $N/2$ recordings. Experimental results are compared with model predictions in high and low correlation regimes.\\
{\em{Spike correlations}--}
To quantify the temporal spike cross correlations between neuron $1$ and $2$ we used the conditional firing rate $\nu_{cond}(\tau)$, which is the firing rate of neuron 2 triggered on the spikes of neuron 1: 
\begin{gather}
\nu_{cond}(\tau)=(\nu_1\nu_2)^{-1/2}\langle s_1(t) s_2(t+\tau)\rangle~\label{CondDefMean}
\end{gather}
This quantity measures correlations on all time scales and is equivalent to the spike count correlation coefficient $\rho$ ($\rho$$\approx$$T(\nu_{cond}(0)$$-$$\nu)$) for small time bin $T$. For a pair of identical neurons $\nu_{cond}(\tau)$ is a symmetrical function which approaches $\nu$ as $\tau$ increases and maximally deviates from $\nu$ at $\tau$$=$$0$. We obtain $\nu_{cond}(0)$ by solving the Gaussian integrals in Eq.~\ref{DefSpikeMeasure},\ref{DefCorCrossInhom} for any $r$:
\begin{align}
\nu_{cond}(0)=\tilde{\nu}\Big(\frac{\nu}{\tilde{\nu}}\Big)^{R}[1+2 r \arctan(\sqrt{R^{-1}})/\sqrt{1-r^2}],\label{Synch}
\end{align}
\begin{figure}[b]
\begin{minipage}{0.47\linewidth}
\includegraphics[width=\linewidth]{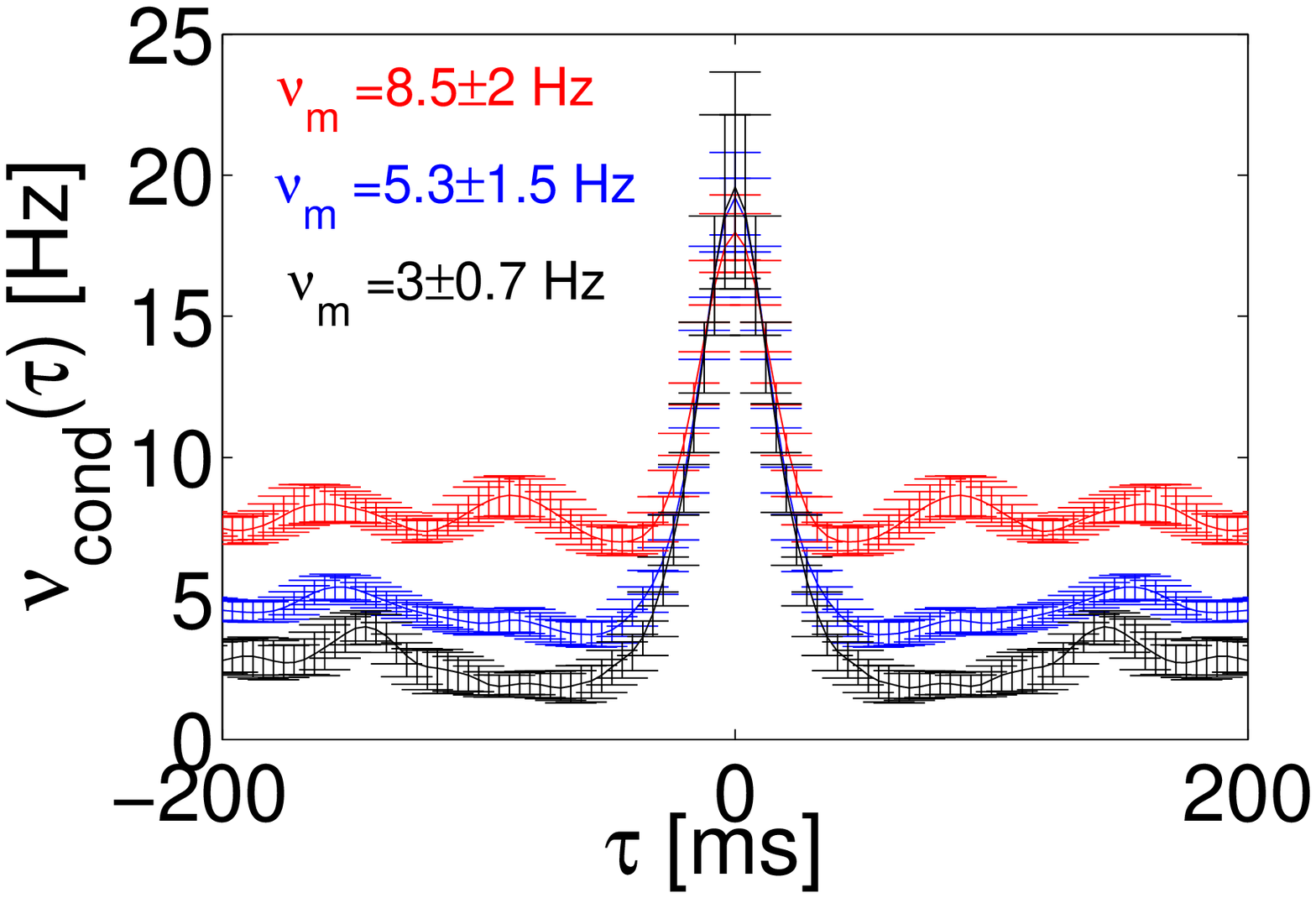}
\end{minipage}
\begin{minipage}{0.47\linewidth}
\includegraphics[width=\linewidth,clip]{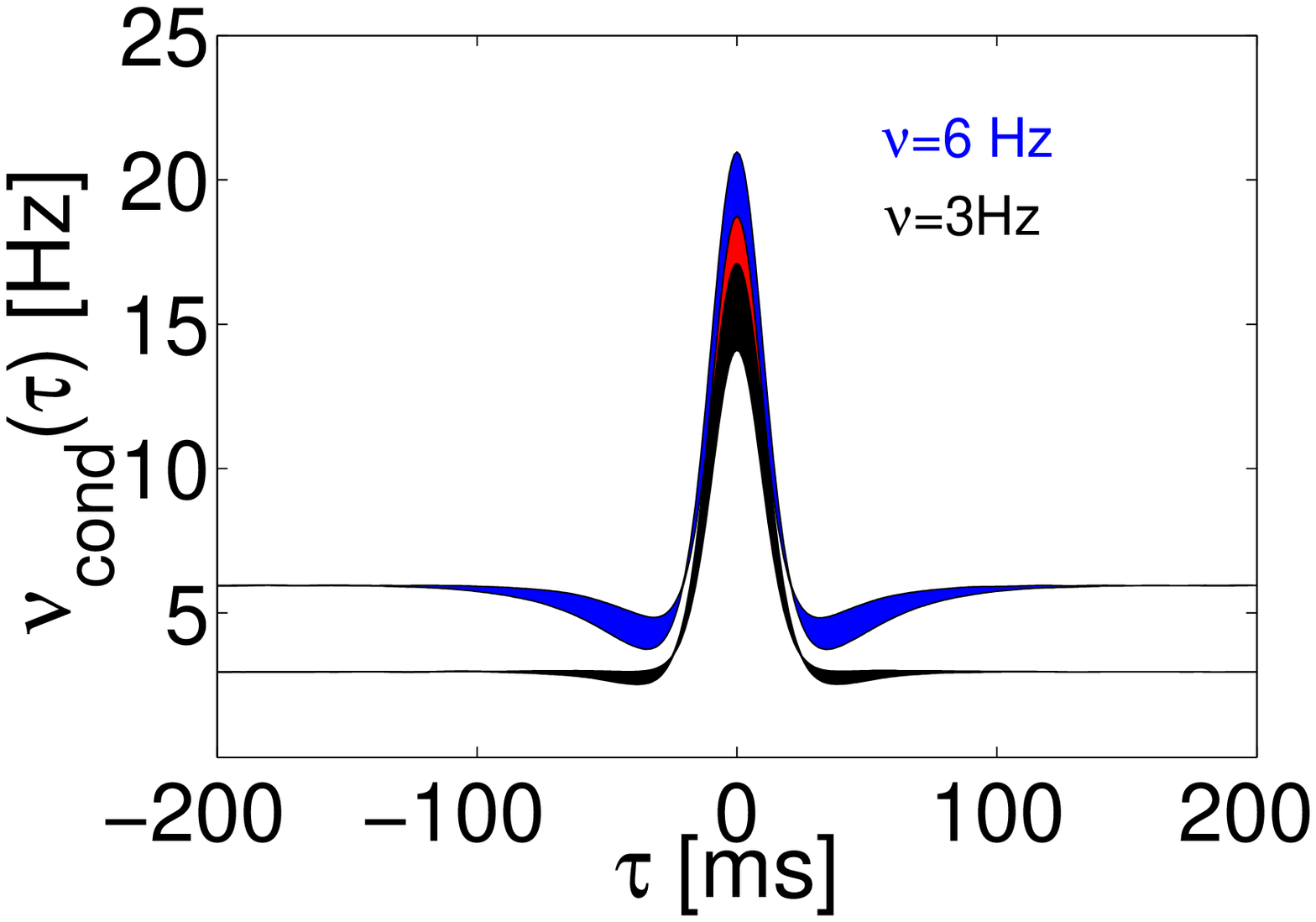} 
\end{minipage}
\caption{Firing rate independence of spike correlations for strong input correlations, theory and experiment. (left) $\nu_{cond}(\tau)$ vs. $\tau$ calculated for identical current injections in recorded neurons. (right) $\nu_{cond}(\tau)$ for simulated pair of neurons for $\nu_1=3\hertz$ (black) and $\nu_2$$=$$6\hertz$ (blue), $12.5\leq\sqrt{1-r}\tau_s$$\leq$$12.8$ with $20\leq$$\tau_s$$\leq25\milli\second$ and $0.59$$\leq$$r$$\leq$$0.75$. The overlap of both families of curves is plotted in red. }
\label{FigLargeCross}
\end{figure}
\noindent where $R=(1-r)/(1+r)$. Eq.~\ref{Synch} predicts a superlinear increase of $\nu_{cond}(0)$ with $r$, see Fig.~\ref{RateDepSmallR}(A). \\
{\em{Strong cross correlations $r$$\approx$$1$}--} In this limit we find that the peak spike correlations are independent of $\nu$ and of the particular shape of $C_{12}(\tau)$. 
\begin{align}
\nu_{cond}(0)&\approx 1/(2\sqrt{2}\sqrt{1-r}\tau_s)~~~~~~~~~~~~~~~~~~(r\approx 1)
\end{align}
Do these predictions hold for neuronal spike correlations? Fig.~\ref{FigLargeCross} (left) depicts recorded $\nu_{cond}(\tau)$ in the high $r$ regime for different firing rates. The correlation peaks for $3,5.3$ and $8.5\hertz$ are essentially identical (Fig.~\ref{FigLargeCross} (left)). These recordings confirm the prediction that the amplitude is insensitive to the firing rate. Additionally, the peak form suggests that there is a rate independent universal correlation peak shape. To assess this possibility theoretically, we calculate $\nu_{cond}(\tau)$ by solving the Gaussian integrals in Eq.~\ref{DefSpikeMeasure},\ref{DefCorCrossInhom} for $r\approx 1$ and $\tau\ll \tau_s$. We obtain:
\begin{align}
\nu_{cond}(\tau)=1/(2\tau_s^*)(1-3/2\hat{\tau}^2+30/16\hat{\tau}^4)+O(\tau^6)\label{bigR}
\end{align}
\begin{figure}[b]
\begin{minipage}[t]{0.48\linewidth}
\includegraphics[width=\linewidth,clip]{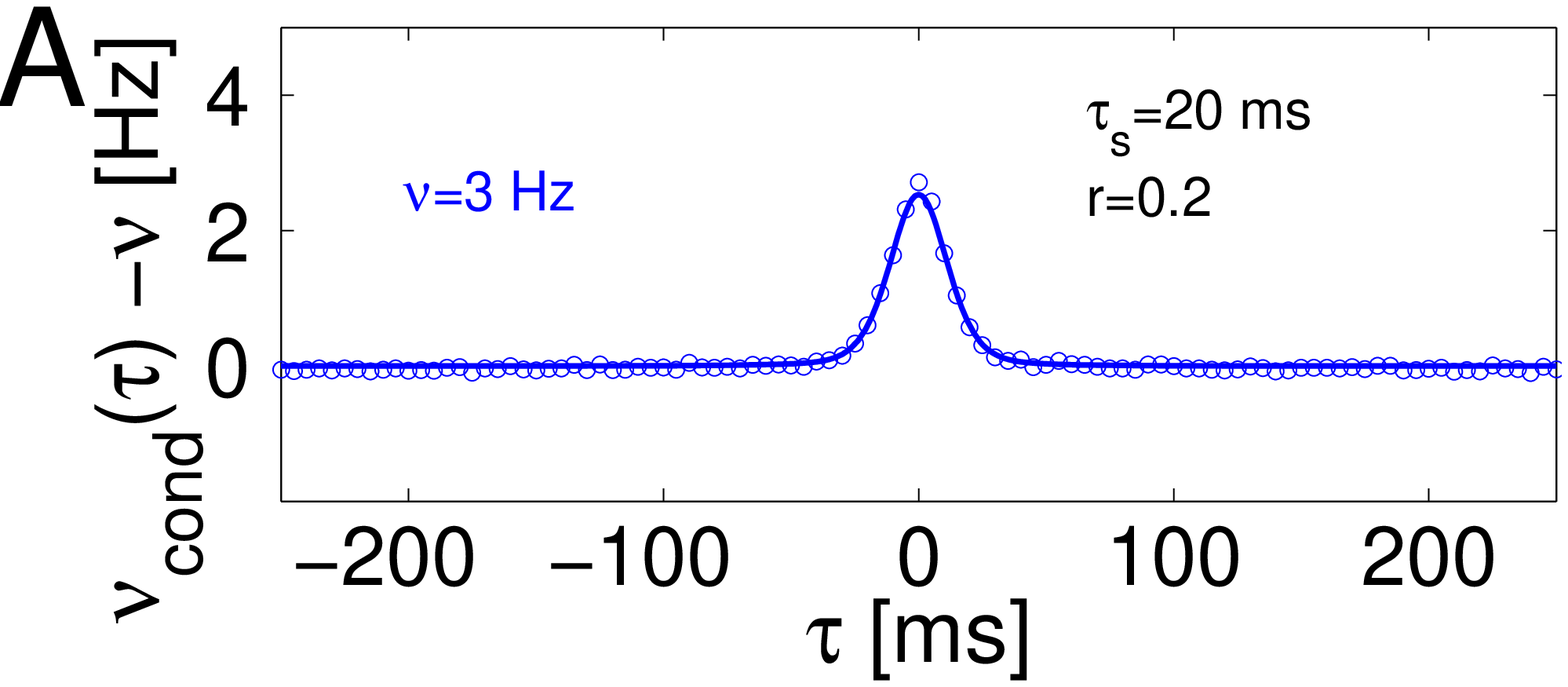}
\end{minipage}
\begin{minipage}[t]{0.48\linewidth}
\includegraphics[width=\linewidth,clip]{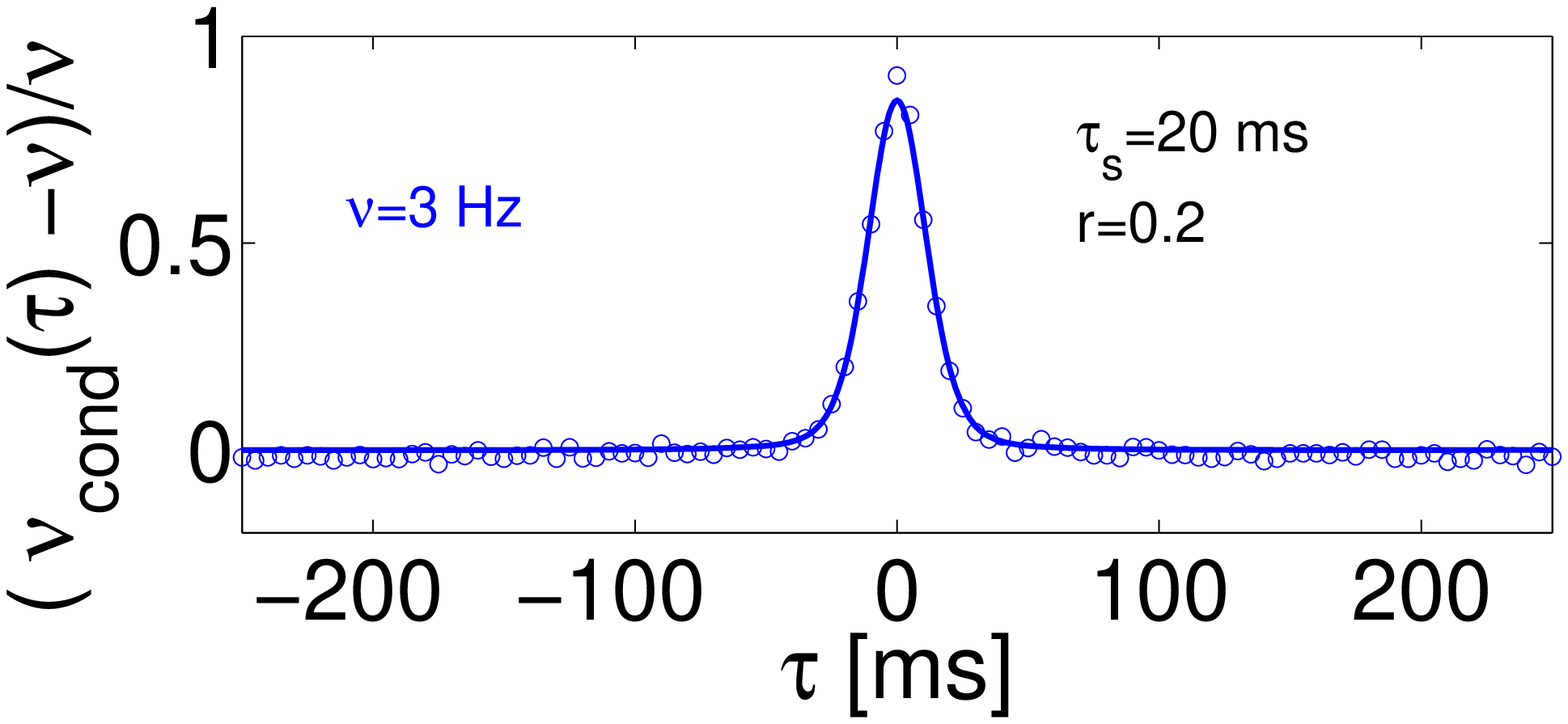}
\end{minipage}
\begin{minipage}[t]{0.48\linewidth}
\includegraphics[width=\linewidth,clip]{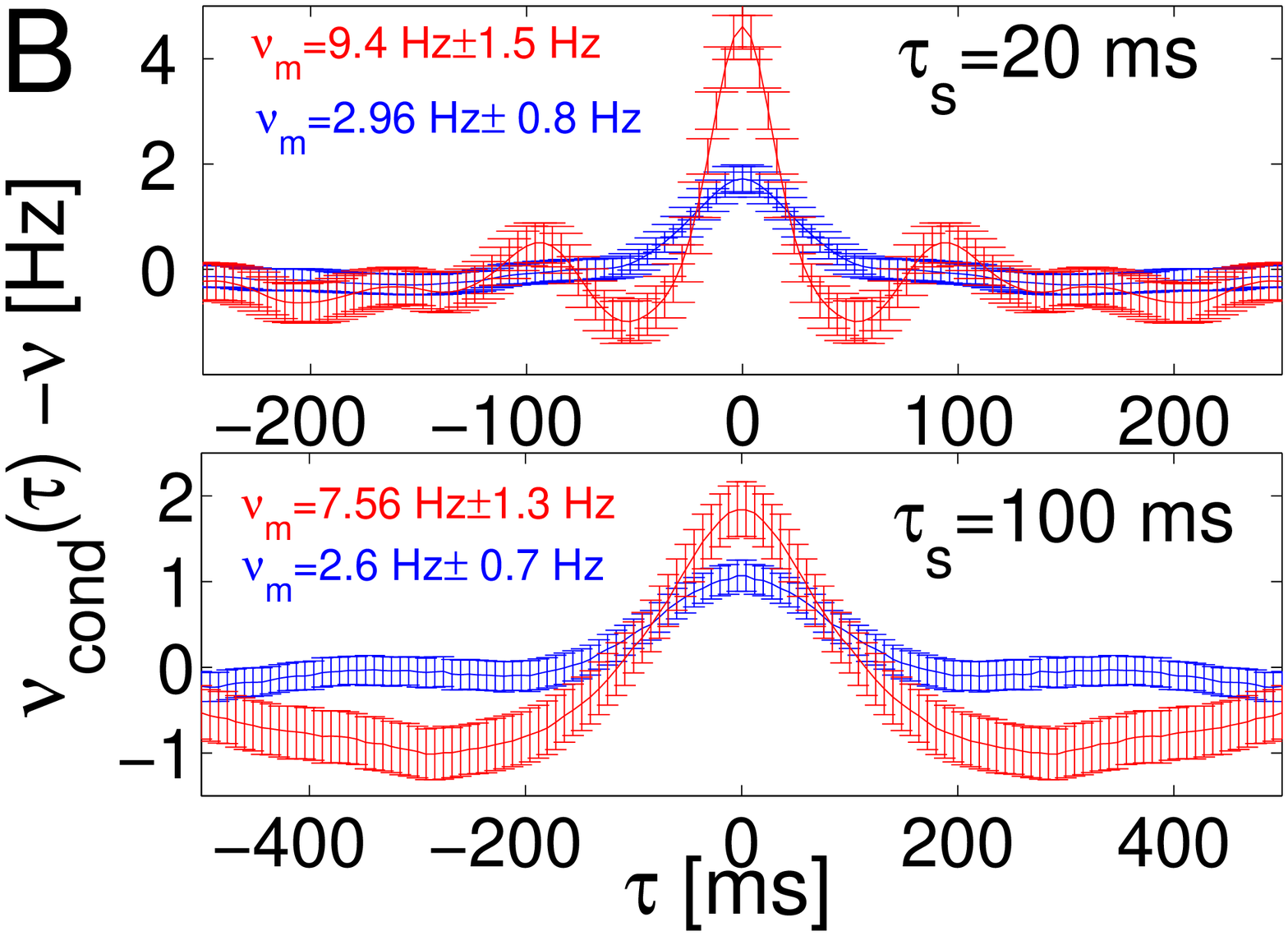}
\end{minipage}
\begin{minipage}[t]{0.48\linewidth}
\includegraphics[width=\linewidth,clip]{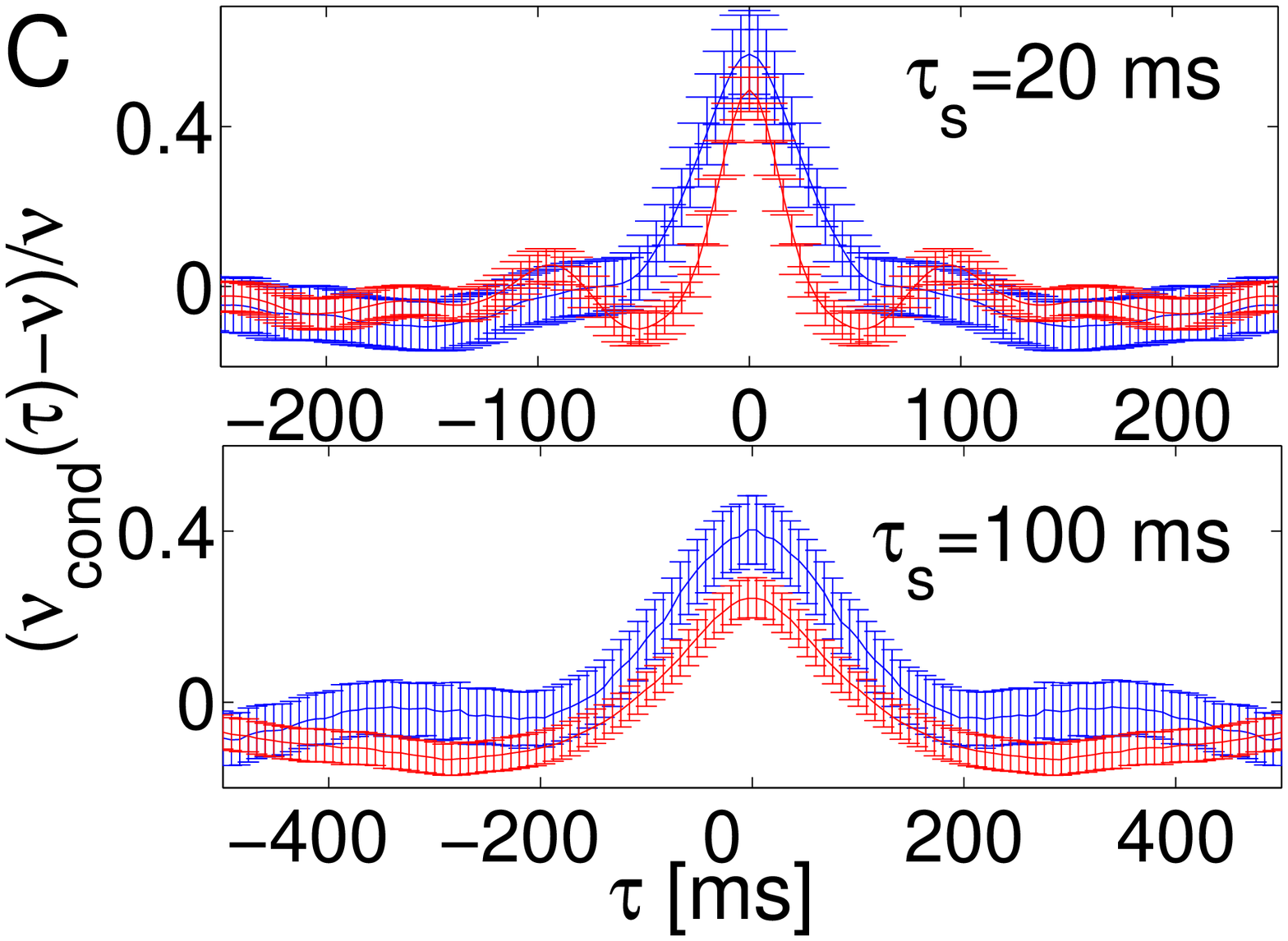}
\end{minipage}
\caption{Firing rate dependence of spike correlations for small $r$, theory and experiment. (A) $\nu_{cond}(\tau)-\nu$ vs $\tau$ and $(\nu_{cond}(\tau)-\nu)/\nu$ vs $\tau$ for $\nu=3\hertz,\tau_s=20\milli\second$; simulation results are superimposed. (B,C) measured $\nu_{cond}(\tau)-\nu$ vs. $\tau$ and $(\nu_{cond}(\tau)-\nu)/\nu$ vs. $\tau$ for $\tau_s\in \{20,100\}\milli\second$; computed from pairs of patch clamp recordings.}
\label{GainExperiments}
\end{figure}
\noindent where $\hat{\tau}$$=$$\tau/\tau^*$ with the time constant $\tau_s^*$=$\sqrt{2}\sqrt{1-r}\tau_s$. For $r=1$ the $\delta$-peak present at the origin of the auto conditional firing rate is recovered. Eq.~\ref{bigR} indeed demonstates the existence of a rate independent universal peak shape and height, determined by $\tau_s$. Note, that Eq.~\ref{bigR} predicts that correlation peak shape is insensitive to the functional form of $C_{12}(\tau)$. 
To explore how close the agreement between theory and experiment is, we computed $\nu_{cond}(\tau)$ from simulated pairs and found a good qualitative agreement with experimental findings for a broad range of parameters. The salient correlation peak structure can be faithfully reproduced by our framework and the additional weak periodic modulation of the experimental stationary rates might be due to additional ion conductances of real neurons. The width and hight of the common peak can be qualitatively described by the theoretical curves in Fig.~\ref{FigLargeCross} (right). We note, that $r$$\approx$$0.7$ in the simulations and in the experiments ($0.6$$<$$r$$<0.9$) were similar and both were close to the typical correlation coefficient ($\rho\approx0.6$ for time bin 40ms) reported recently for cell pairs subject to identical white noise currents [Fig.1d in~\cite{Rocha}].\\
{\em{Low correlation strength--}} In this limit, $\nu_{cond}(\tau)$ recovers the rate dependence and shows sensitivity to $c(\tau)$. This is valid for $rc(\tau)\ll 1$. We obtain $\nu_{cond}(\tau)=\nu+rg(\tau)+O(r^2)$ by solving the Gaussian integrals in Eq.~\ref{DefSpikeMeasure}, \ref{DefCorCrossInhom}. Here, $g(\tau)$:
\begin{align}
g(\tau)&=\nu(c(\tau)|2\log(\nu/\tilde{\nu})|-\pi/2\tau_s^2c''(\tau))\label{SmallRHom}\\
g(0)&=\nu(|2\log(\nu/\tilde{\nu})|+\pi/2)\label{LinCor}
\end{align}
Fig.~\ref{RateDepSmallR}(C) illustrates the dependence of $\nu_{cond}(\tau)$ on $\tau_s$ and $\nu$. Eq.~\ref{SmallRHom} implies that the spikes are typically correlated on a shorter time scale than the underlying MPs, due to the admixture of $c''(\tau)$ which has a shorter time scale than $c(\tau)$. For low rates, the contribution of $c''(\tau)$ is negligible, but already at $20\%$ of $\tilde{\nu}$ the influence of $c''(\tau)$ cannot be neglected leading to a sharpening of spike correlations. Notably, the spike correlations with temporal widths much smaller than the underlying voltage correlations have been previously observed {\em{in vivo}}[p. 367 in ~\cite{Lampl}]. Eq.~\ref{LinCor} also implies an increase of correlation of spikes  with $\nu$ (Fig.~\ref{RateDepSmallR} B). However, the percentage of simultaneous spikes ($(\nu_{cond}(\tau)-\nu)/\nu$) is higher for lower rates, because at low firing rates ($\psi_0$$\gg$$\sigma_V$) the additional common component $n_c(t)$ is more critical for reaching the threshold, than $n_i(t)$. $g(0)$ attains a maximal value $g_{max}$$=$$2\nu_{max}$ for the rate $\nu_{max}$$=$$(2\pi\tau_s)^{-1}\exp(\pi/4-1)$ (Fig.~\ref{RateDepSmallR},A). This simple model qualitatively captures the salient correlation peak for low firing rates recorded in neurons subject to weak correlated input(Fig.~\ref{RateDepSmallR}(C)), the presence of weak damped periodic modulation might be due to additional ion conductances of real neurons. The $\nu$-dependence predicted by Eq.~\ref{LinCor} is also evident in Fig.~\ref{GainExperiments} (B), even though $\nu_M=9.4\hertz$ escapes direct comparison ($\nu_m$$>$$\tilde{\nu}$): with increasing $\nu$, $\nu_{cond}(0)$$-$$\nu$ increases and  $(\nu_{cond}(0)-\nu)/\nu$ is decreasing with $\nu$. Both results are consistent with recent reports (Fig.1c in~\cite{Rocha} and Fig.3(B,C) in~\cite{Svirskis}).\\
\begin{figure}[t]
\begin{minipage}{0.47\linewidth}
\includegraphics[width=\linewidth]{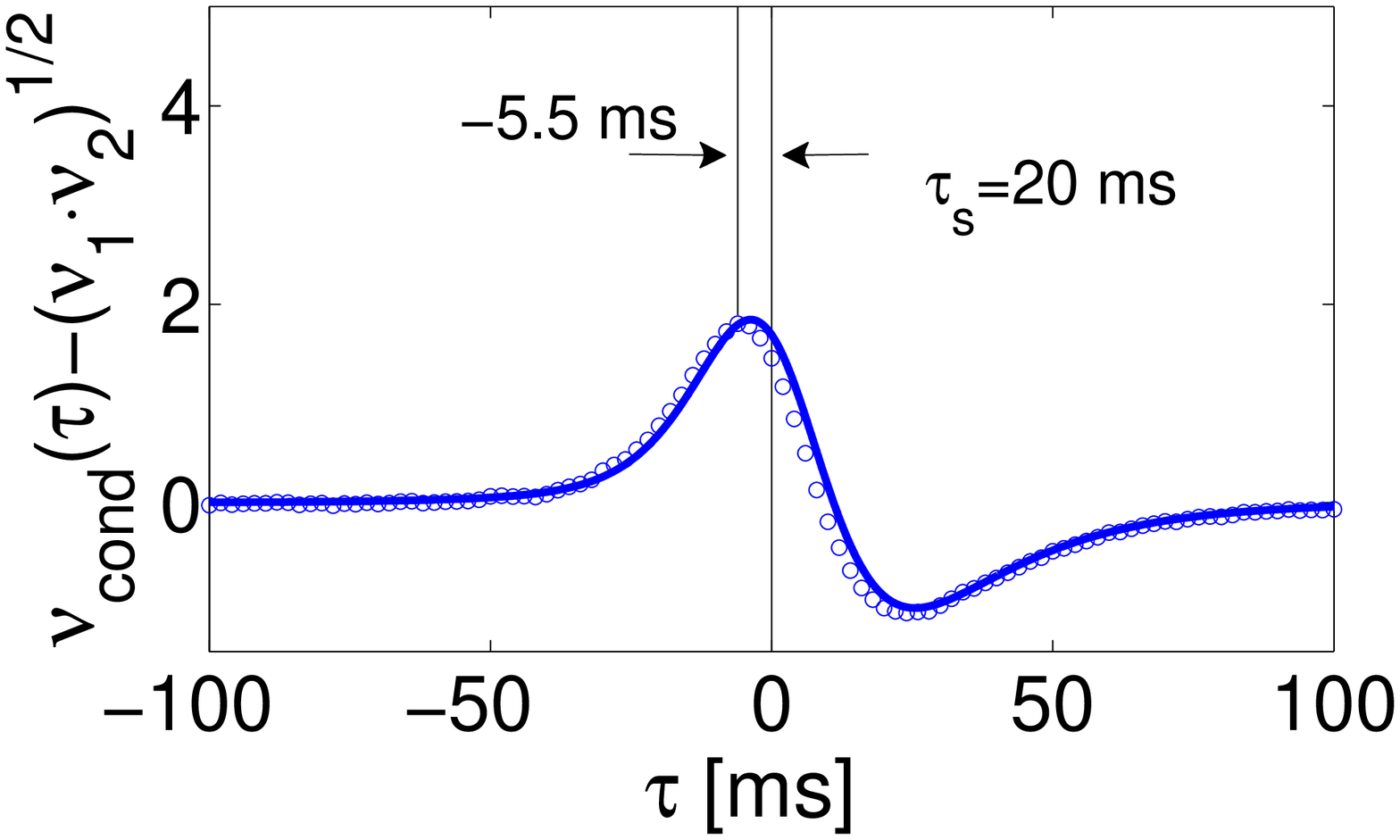}
\end{minipage}
\begin{minipage}{0.47\linewidth}
\includegraphics[width=\linewidth]{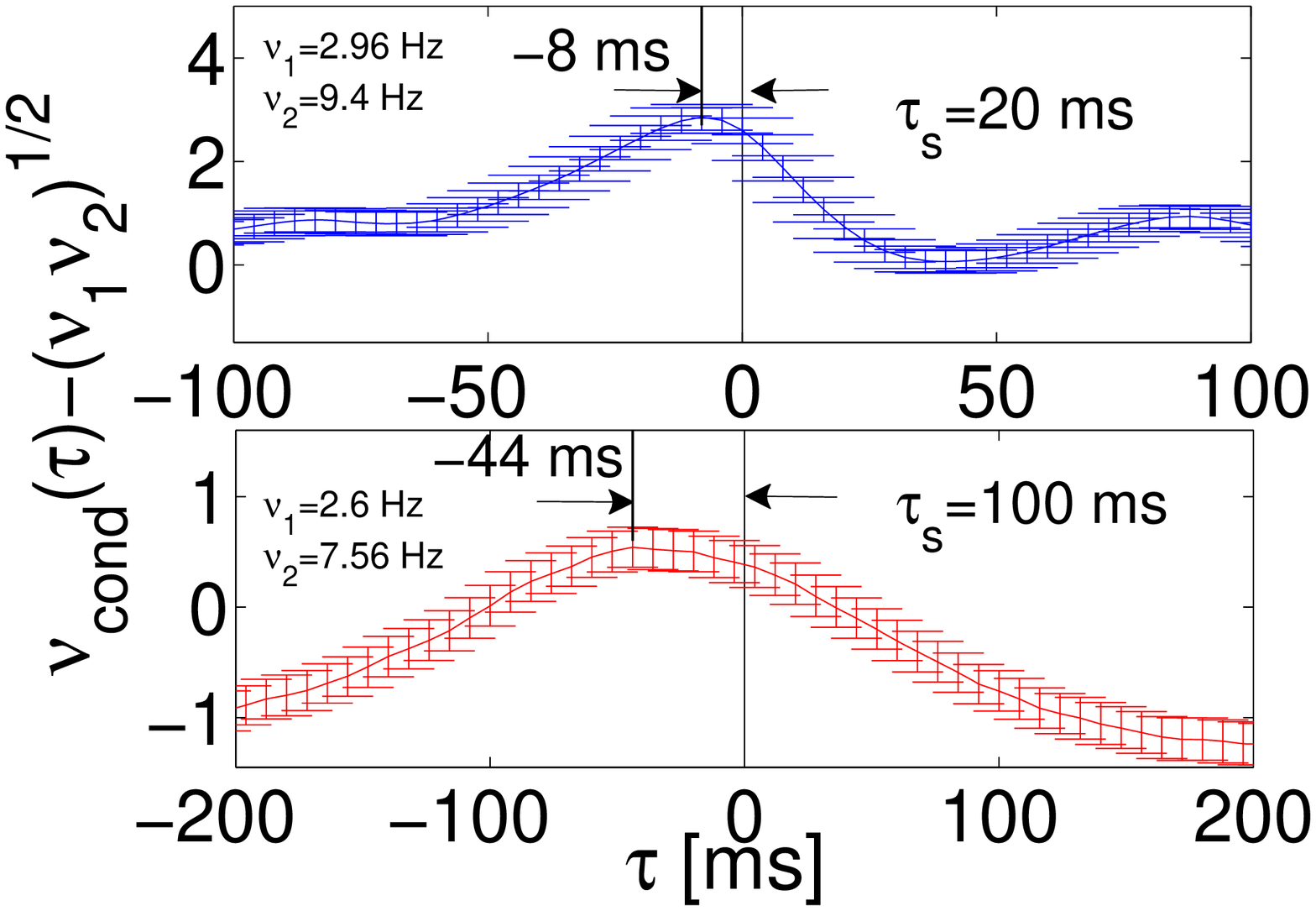}
\end{minipage}
\caption{Difference in firing rate leads to asymmetric spike correlations. (left) $\nu_{cond}(\tau)-\sqrt{\nu_1\nu_2}$ vs. $\tau$, simulation results and theoretical second order prediction for $r$=$0.2$, $\nu_1$=$2.65\hertz$, $\nu_2$=$3\cdot2.65\hertz$ and $\tau_s$=$20\milli\second$. Arrows denote the peak position. (right) $\nu_{cond}(\tau)-\sqrt{\nu_1\nu_2}$ vs. $\tau$. Experimental results for $\tau_s=20,100\milli\second$ for different rates.}
\label{ShiftPeak}
\end{figure}
{\em{Rate asymmetry--}}
The correlation matrix in Eq.~\ref{DefCorCrossInhom} includes $c'(\tau)$, which so far did not enter $\nu_{cond}(\tau)$. As $c'(\tau)$ is an antisymmetric function it is conceivable to assume that a broken symmetry  ($\nu_1\neq\nu_2$) will lead to asymmetric $\nu_{cond}(\tau)$. In the low $r$ regime, we obtain $g(\tau)$=$(\nu_{cond}(\tau)-\sqrt{\nu_1\nu_2})/r$ by solving the Gaussian integrals in Eq.~\ref{DefSpikeMeasure},\ref{DefCorCrossInhom}. $g(\tau)$ then is:
\begin{align}
g(\tau)&=\sqrt{\nu_1\nu_2}(c(\tau)e_1e_2-\pi/2\tau_s^2 c''(\tau)-c'(\tau)\tau_s\Delta)\label{Asym}\\
\tau_{\text{Peak}}&=\tau_s\Delta/(e_1e_2+\pi c^{(4)}(0)\tau_s^4/2).\label{tauPeak}
\end{align}
where $e_i$=$\psi_0/\sigma_{V_i}$ and $\Delta$=$\sqrt{\pi/2}(e_2-e_1)$. The peak position is no longer at $\tau$=$0$ but is shifted to $\tau_{\text{Peak}}$.
$c'(\tau)$ leads to a temporal delay of spikes indicating that the spikes of the higher rate neuron precede spikes of the lower rate neuron, despite the perfect synchrony of the common input. The asymmetry increases with increasing difference of threshold-to-variance ratios and increasing $\tau_s$. Theoretically predicted asymmetric $\nu_{cond}(\tau)$ (Fig.~\ref{ShiftPeak} (left)) is in good agreement with experimental results (Fig.~\ref{ShiftPeak} (top right)). Measured $\nu_{cond}(\tau)$ also shows an increase of the temporal shift with $\tau_s$. This $\tau_s$ dependence is in qualitative agreement with Eq.~\ref{tauPeak}, despite the fact that experiments with $\tau_s$$=$$100\milli\second$ escape direct quantitative comparison as $\nu_m$$>$$\tilde{\nu}$$\approx$$1.6\hertz$. Notably, shifted correlations are well known in the biological literature and are often interpreted as indications of synaptic connections or the presence of delayed inputs~\cite{Schneider}. However, our framework reveals a potential mechanism for the occurrence of asymmetric correlations in pairs with synchronous inputs~\cite{Lampl}.\\
{\em{Discussion--}} We presented a framework for the description of the auto- and cross correlations of upward level crossings with arbitrary functional form of input correlations. Our results confirm previous reports on the rate dependence of spike correlations~\cite{Rocha,Svirskis}. This behavior, however, holds for weak correlations only. With strongly correlated inputs, spike correlations become independent of the firing rate but depend on the correlation time of voltage fluctuations. In cell pairs with rate differences the temporal symmetry of spike correlations is lost. Finally, let us stress that input correlations modeled in our framework do not imply a particular connectivity as they can arise from common input or reciprocal connections. Identifying self-consistent choices of $c(\tau)$, $r$, $\psi_{0,i}$, $\sigma_{V,i}$ in a network of prescribed connectivity will be a fruitful direction of future research.\\ 
{\em{Acknowledgments}} We thank E. Nikitin for help conducting the experiments, I. Fleidervish, M.Gutnick, A. Witt, M. Huang, W. Wei, B. Kriener, W. Keil for fruitful discussions and A. Witt for help with statistical analysis. We thank BMBF, GIF and Max Planck Society for support.

\end{document}